\documentstyle[11pt,psfig,aaspp4,flushrt]{article}
\def\be{\begin{equation}}
\def\ee{\end{equation}}
\def\etal{{\it et al. }}

\def\micron{$\mu$m}

\def\res{${\cal\char'122}$}
\def\resm{\cal\char'122}
\def\source{${\cal\char'123}$}
\def\sourcem{{\cal\char'123}}

\def\bgm{{\cal\char'102}}
\def\at{${\cal\char'101}$}
\def\atm{{\cal\char'101}}
\def\trs{${\cal\char'124}$}
\def\trsm{{\cal\char'124}}

\def\qu{${\cal\char'121}$}
\def\qum{{\cal\char'121}}
\def\dark{${\cal\char'104}$}
\def\darkm{{\cal\char'104}}
\slugcomment{Version 3.0, Feb. 2001}
\begin{document}

\title{The Optical/Infrared Astronomical Quality of High Atacama Sites.
II. Infrared Characteristics}
\author {\it Riccardo Giovanelli, Jeremy Darling,
Charles Henderson, William Hoffman, Don Barry, James Cordes, 
Stephen Eikenberry, George Gull, Luke Keller, J.D. Smith \& Gordon Stacey}
\affil{$^1$Department of Astronomy, Cornell University, Ithaca, NY 14853}

\hsize 6.5 truein

\begin {abstract}
We discuss properties of the atmospheric water vapor above the high Andean 
plateau region known as the {\it Llano de Chajnantor}, in the Atacama Desert 
of northern Chile. A combination of radiometric and radiosonde measurements 
indicate that the median column of precipitable water vapor (PWV) above the 
plateau at an elevation of 5000 m is approximately 1.2 mm. The exponential
scaleheight of the water vapor density in the median Chajnantor atmosphere 
is 1.13 km; the median PWV is 0.5 mm above an elevation of 5750 m. Both of
these numbers appear to be lower at night. Annual, diurnal and 
other dependences of PWV and its scaleheight are discussed, as well as the 
occurrence of temperature inversion layers below the elevation of peaks 
surrounding the plateau. We estimate the background for infrared observations
and sensitivities for broad band and high resolution spectroscopy. The results 
suggest that exceptional atmospheric conditions are present in the region, yielding 
high infrared transparency and high sensitivity for future ground--based infrared 
telescopes. 

\end{abstract}

\noindent {\bf Subject Headings}: Astronomical instrumentation, methods and techniques:
atmospheric effects, site testing.

\section {Introduction}

Atmospheric molecules are responsible for band absorption of cosmic radiation.
In the near and mid--IR, the most important absorbers are the triatomic 
molecules H$_2$O, CO$_2$ and O$_3$. 
Between absorption bands, partially transparent 
windows appear, making astronomical observations possible in the near and 
mid--IR from the ground. The atmosphere is virtually opaque between 50 and 
200 \micron ~from all sites on the ground, making that wavelength region
the exclusive territory of airborne and space observatories. Water vapor 
also dominates absorption in the far IR and submm region; atmospheric 
windows appear near 200, 350, 450, 600, 750 and 870 \micron; in this spectral 
region transparency is significant only in dry, high altitude sites 
(Tokunaga 1998; Wolfe \& Zissis 1978). 
H$_2$O molecules have a short residence time (a few days) in the atmosphere and 
their concentration is highly variable. Thus, the characterization 
of potential astronomical sites for operation in the IR and submm spectral 
regimes depends very strongly on measurements of atmospheric water vapor. 

In a companion paper (Giovanelli \etal 2000; Paper I), we have discussed the 
desirability of ascertaining the quality of high altitude sites for astronomical
research in the optical and infrared parts of the spectrum. Paper I describes 
briefly the Atacama desert region and, as part of a site survey campaign, the
results of optical seeing measurements. Here, we review the results of 
measurements carried out in the Andean plateau known as the Llano de Chajnantor,
in the Atacama desert of northern Chile, aimed at gauging the atmospheric 
transparency and emissivity in the infrared. The main characteristics of the 
atmospheric water vapor in the region are discussed: its absolute 
amount, annual, seasonal and diurnal variations, as well as its vertical 
distribution. The latter information provides indications of the relative 
infrared transparency at sites at varying elevation above the plateau, even of 
those currently inaccessible for deployment of testing equipment. 

In Section \ref{data}~ we briefly describe the data utilized for the 
analysis; in Section \ref{pwv_tot}, statistics of the total precipitable water 
vapor (PWV, defined as the atmospheric column density of water vapor) are given, 
and in Section \ref{pwv_vert} ~the vertical distribution of PWV is analyzed, 
with estimates of its value expected for candidate sites for an optical/IR 
telescope. Finally, in Section \ref{transp} ~we estimate atmospheric transparency, 
emissivity and telescope sensitivities in the infrared, for a range of possible 
water vapor conditions.

\section {Data Sources \label{data}}

We make use of two kinds of data in this report: 
\begin{itemize}
\item Atmospheric opacities measured with automated  tipping radiometers at 
225 GHz and at 183 GHz. These instruments are operated by
the National Radio Astronomy Observatory (NRAO) , the European Southern
Observatory (ESO) and the Onsala Space Observatory. Data taken since April 
1995 can be found at {\it http://www.tuc.nrao.edu/mma/sites/Chajnantor}
and {\it http://alma.sc.eso.org}. We shall refer to summaries of relevant 
data products as given at those sites.
\item Radiosonde data from balloon launches initiated in October 1998, jointly 
operated by Cornell University, NRAO, ESO and the Smithsonian Astrophysical 
Observatory. We will summarize results from radiosondes launched between October 
1998 and August 2000. Details on individual launches can be found at
{\it http://www.astro.cornell.edu/atacama/sondedata.html} and at the 
above web sites of NRAO (Radford 2000) and ESO (Otarola 2000). A more 
detailed analysis of the full radiosonde data set will be presented in a 
forthcoming paper, by the `radiosonde consortium'.
\end{itemize}
Tipping radiometers yield values of the atmospheric opacity, from which 
estimates of PWV above the radiometer location can be inferred. Radiosondes, 
on the other hand, provide vertical profiles of atmospheric parameters 
(temperature, pressure, relative humidity, wind speed and direction); 
from those, the water vapor density distribution and thus the PWV above 
any altitude can be inferred. Several summits, which are potentially 
attractive sites for infrared telescopes, have no access roads and thus 
installation and access to equipment at those locations is impractical. 
The free atmosphere parameters obtained from radiosonde launches provide 
a rough approximation of conditions that would be found at those summits.
In the remainder of this paper, we shall always refer to PWV at zenith. 

The sites for radiometer data taking and radiosonde launches are within one km
from each other, near latitude S 23$^\circ$ 01', longitude W 67$^\circ$ 46',
at an altitude near 5000 m above mean sea level. It should be 
kept in mind that, depending on wind speed and direction,  a sonde flight 
samples an oblique path through the atmosphere, rather than a vertical one. 
In strong wind. a sonde profile typically spans tens of km in horizontal range. 
However, since most of the atmospheric water vapor is generally distributed 
within a few km from the ground, the effective sampling of most of the PWV 
corresponds to a horizontal range of only a few km from the launch site. Given 
the prevalent wind direction (see Paper I for details), sondes usually fly 
towards the East over unpopulated terrain, with ground elevation ranging 
between 4500 m and 5800 m above mean sea level.

In this paper, we shall use data from 108 radiosonde launches distributed as
follows: 8, 10 and 16 launches in October, November and December 1998 
respectively; 1, 5 and 33 in February, March and November 1999 respectively; 
11, 18 and 6 in June, July and August 2000 respectively. Thirty were launched
in nighttime hours (UT 01 to 11 hours), 65 in daytime (UT 12 to 21 hours) 
and 13 in the early evening (UT 22 to 00 hours). For each sonde profile we 
integrate the water vapor to obtain PWV above the plateau level of 5050 m, 
PWV above the elevation of 5400 m (equivalent to the summit of Cerro Honar) 
and PWV above 5750 m (equivalent to the summit of Cerro Chasc\' on). We  
also compute the height above the plateau at which water vapor density 
drops to $1/e$ of the value at the plateau level, $h_e$, an indication of 
the exponential scaleheight of the water vapor layer. Since the water vapor 
distribution is most often not close to an exponential, we also measure 
($h_{1/2}$), the height at which PWV drops to 50\% of the value at the 
plateau level. Finally, we note the location of temperature inversions, 
which often are seen trapping large fractions of the total PWV below them. 
The derivation of these parameters and a detailed presentation of the sonde 
data will be given in a `consortium paper', as mentioned above.

\section {Precipitable Water Vapor Above 5000 m \label{pwv_tot}}

The conversion from radio opacities to PWV can be obtained from models of the
atmosphere and assumptions on the water vapor line shapes. These are usually
fairly uncertain, especially in the case of 225 GHz observations (e.g. see 
Holdaway \etal 1996). The conversion is more reliable in the case of 183 GHz
data, for the nearby water line is stronger and the ``wet'' contribution to the 
opacity larger. Delgado \etal (1999) give a reliable conversion relation
directly from antenna temperature to PWV from 183 GHz observations, and have 
also derived a conversion relation from $\tau_{225}$ to PWV, which we use:
\be
\tau_{225} = 0.0435 (PWV) + 0.0068  \label{taupwv}
\ee
Based on preliminary comparisons, Equation \ref{taupwv}~ yields good agreement 
with PWV derived from radiosonde data for values of PWV$<3$ mm; for higher 
values, however, PWV may be overestimated by Eqn. \ref{taupwv}. Fortunately, 
the latter cases are of relatively less interest in our case. 

The most extensive data base of radiometric opacities at the Chajnantor Plateau 
is that derived from the NRAO data base (Radford 2000). The historical quartiles 
between April 1995 and October 2000 are respectively 0.036, 0.060 and 0.114,
which convert to PWV values of respectively 0.67, 1.22 and 2.46 mm. 
The PWV quartiles of 108 radiosonde flights between October 1998 and August 2000 
are: 0.71, 1.04 and 1.85 mm. The difference between radiosonde and tipping
radiometry values of PWV is not very surprising, for variations in the median
PWV from year to year are large and the time interval for sonde launches
is less than two years. Moreover, the radiosonde launches underemphasize the
Summer season, when PWV is much larger than in the rest of the year. 
The value of PWV inferred from radiometry provides
a more reliable long--term estimate for the Chajnantor Plateau, although it
should be kept in mind that the epoch of radiometry measurements includes the 
exceptional {\it El Ni\~ no}--{\it La Ni\~ na} episode of 1997--98, the 
most extreme on record (McPhaden 1999), which may bias high the PWV median.

Seasonal variations in PWV are conspicuous. In the months of January and February, 
PWV is several times higher than in the winter months. This is the result of moist
Amazon air flowing from the NE, a condition locally referred to as ``Bolivian 
Winter'' and often accompanied by precipitation above 4000 m elevation. Inspection 
of the 225 GHz radiometry record indicates that the median PWV at Chajnantor between 
March and early December is less than 1 mm.

\begin{figure}
\plotone{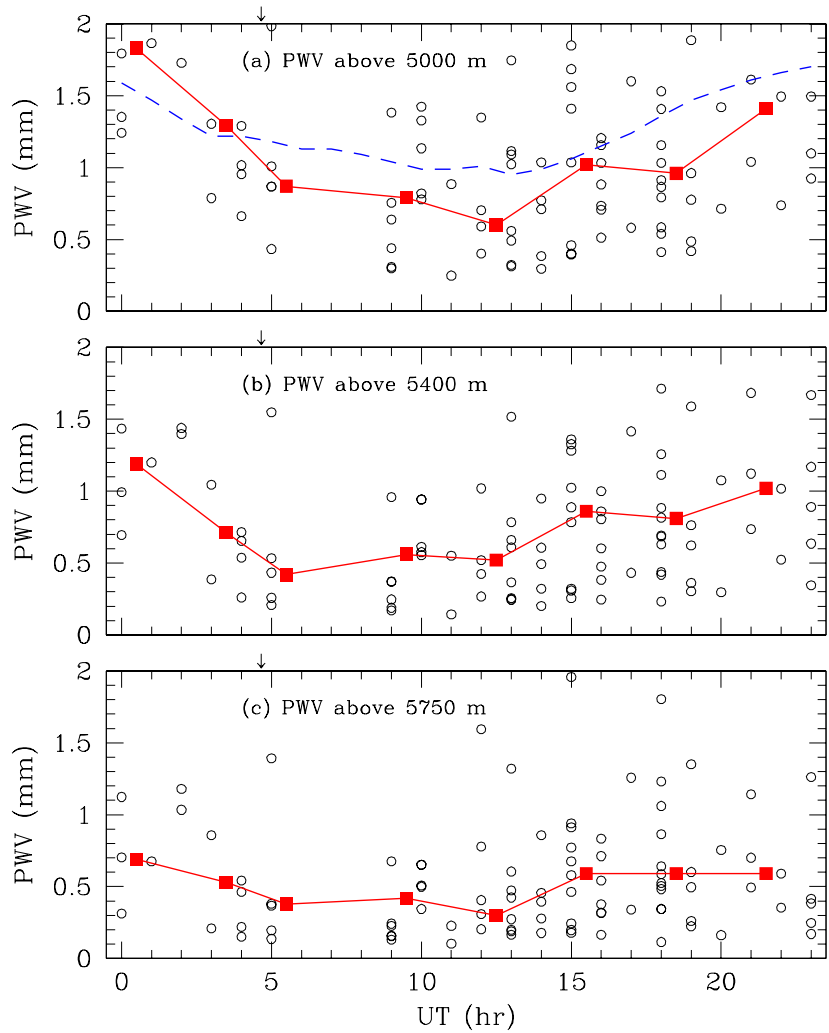}
\caption{PWV as obtained from radiosonde profiles. The water vapor density 
is integrated repsectively above 5000 m, above 5400 m and above 5750 m. 
Unfilled circles refer to individual sonde profiles, filled squares are 
running medians over 3 hr intervals. Midnight is at 4:31 h UT, as 
indicated by the vertical arrow on top of each panel.
\label{pwv_ut}}
\end{figure}

The diurnal cycle also affects PWV, which tends to be higher in the afternoon 
hours, followed by a rapid decrease after sunset. Figure \ref{pwv_ut} ~displays 
the diurnal variation of PWV above 5000 m, above 5400 m and above 5750 m, 
respectively top to bottom, as derived from radiosonde data (unfilled circles). 
Median values over 3 hr intervals are indicated as large, filled
squares, connected by a solid line. The dashed line displays the median values
of PWV, as obtained from 225 GHz opacities at 5000 m. Local midnight takes place
at 04:31 UT. The atmosphere appears to be driest in the late part of the night
and early part of the morning. The diurnal cycle of PWV lags in phase the 
sunlight cycle by about 4 hours, and has amplitude of about 20\% about the 
median value.

\section {Vertical Distribution of Precipitable Water Vapor \label{pwv_vert}}

Table \ref{pwvtab} ~displays quartile values of PWV above 5000, 5400 and 
5750 m for four different groupings of the radiosonde flights: in addition 
to the daytime and nighttime sets described in Section \ref{data}, we 
also use the combined set of 108 sondes (``All'') and that of 32 sondes 
launched between UT 05 and UT 13 hrs. The latter corresponds to the
UT interval in which the minimum of the diurnal cycle is seen. We note
that with respect to the value of PWV at the plateau level overall, 
PWV drops by 30\% in the first 400 m, and by a factor of 2 in the
first 750 m above the plateau. That decrease may be even steeper
during nighttime hours, suggesting that the effective thickness of
the water vapor layer decreases at night. While the absolute values
of PWV listed in Table 1 may be lower than historical values, as
indicated by the comparison with 225 GHz opacities and discussed in
the preceding Section, the variation of PWV with elevation is reliable.  

Whether the decrease in PWV with height will be that of the free atmosphere 
indicated in Table 1, or lower, will depend on the local topography, the 
area of the summits, the direction of the wind and other factors. However,
it is clear that by accessing higher ground above the plateau, significant 
gains in terms of infrared transparency and background emission can be 
obtained. 

\begin{figure}
\plotone{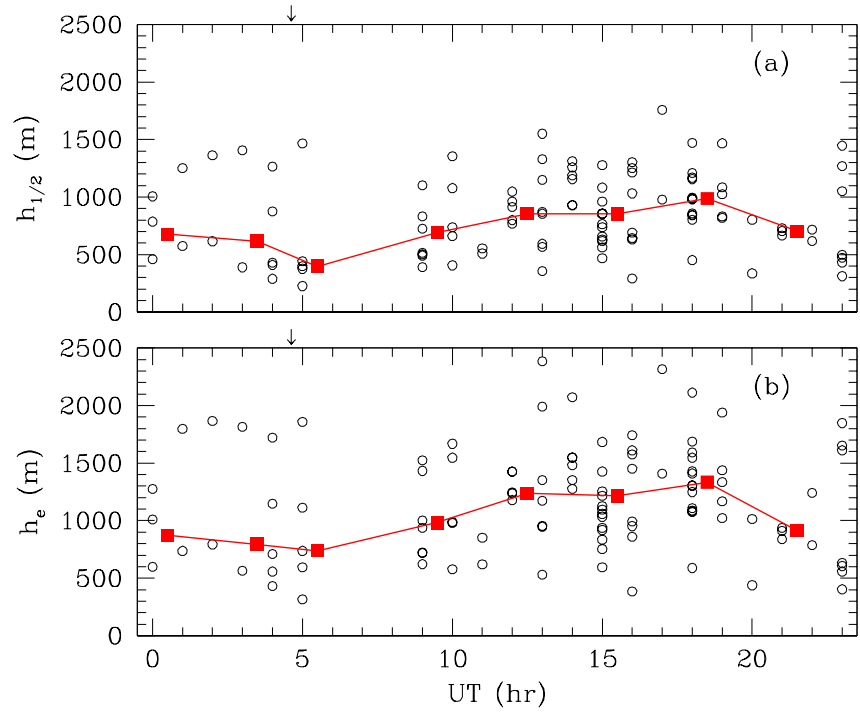}
\caption{Half--thickness and scaleheight of the water vapor layer, as 
obtained from radiosonde profiles. Unfilled circles refer to individual 
sonde profiles, filled squares are running medians over 3 hr intervals. 
Midnight is at 4:31 h UT, as indicated by the vertical arrow on top
of each panel.
\label{h_ut}}
\end{figure}

The thickness of the water vapor layer can be measured by $h_e$ or $h_{1/2}$.
For the set of 108 radiosondes, we find the following quartile values:
\begin{tabbing}
~~~~~~~~~~~~~~~~~~~~~~~~~~~~~~~~~~~\=$h_{1/2}$~\=:~~~ \=571, ~~\=836, ~~\=1083 ~~~ \=m \\
\>$h_e$     \>:    \>836,   \>1135,  \>1504     \>m \\
\end{tabbing}
Examining the relative humidity records from stations at different elevations
at the Japanese testing site of Rio Frio, a region to the south of the
Chajnantor Plateau, Holdaway \etal ~(1996) found indications that the 
scaleheight of the water vapor layer may vary with the diurnal cycle,
between values of approximately 2 and 1 km, the low value being attained
at night. Figure \ref{h_ut}~ yields support for that suggestion. Values of
$h_e$ and $h_{1/2}$ for each in the set of 108 radiosonde flights are
displayed against time of day. Median values are identified by large
filled squares. A diurnal oscillation in the thickness of the water
layer, of amplitude near 25\% about its median value, may be present.
The cycle appears to be in phase with that of the solar illumination cycle.
The statistical significance of this result is low, the lack of sonde
launches between UT 04 and 09 hrs being particularly important in this
respect. We next present corroborating evidence regarding this potentially
important result.

\begin{figure}
\plotone{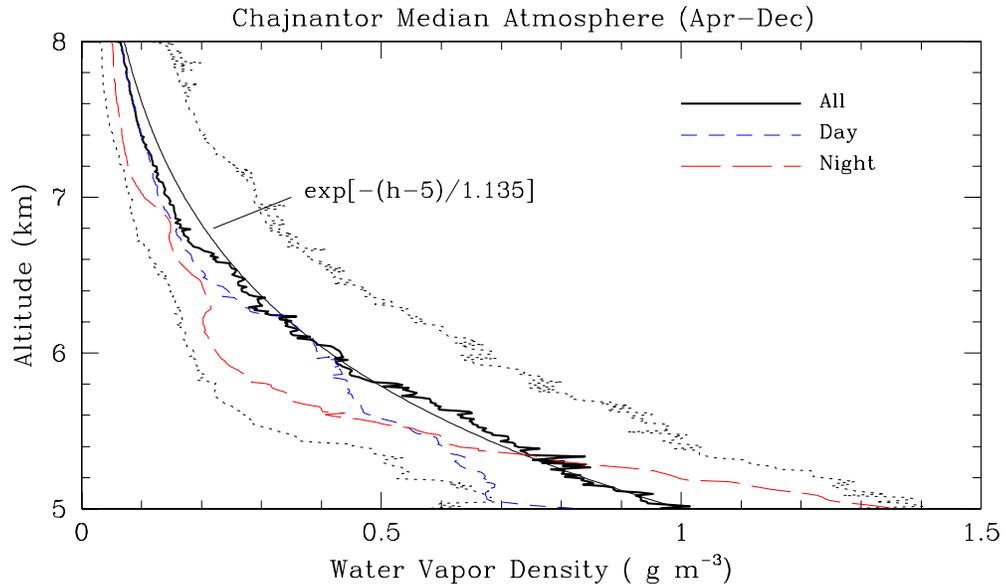}
\caption{Distribution of the water vapor density in the median
atmosphere over the Chajnantor Plateau. The thick, solid line
tracks the median water vapor density, at each altitude, from 108
radiosonde profiles; the dotted lines indicate the 25\% and 75\% 
quartiles about that median. The thin solid line is an exponential
of scaleheight 1.135 km. The short--dash line is the median profile
for 65 daytime sondes, while the long--dash line is the median profile
for 30 nighttime sondes. Note that the number of night plus day sondes
is less than the total, which includes also twilight launches.
\label{medwv}}
\end{figure}

In Figure \ref{medwv} ~we show the vertical distribution of the water
vapor density in the Chajnantor median atmosphere (similar plots for
the temperature distribution and the wind speed are shown in Paper I).
The thick solid line tracks the median water vapor density profile over
the plateau between April and December, from 108 radiosonde profiles.
The distribution is well approximated by an exponential with scaleheight
$h_e = 1.135$ km, as discussed above. The median profiles for day and night 
radiosonde launches do however depart significantly from exponential 
behavior: during night, more water vapor is found at lower elevations,
i.e. the effective thickness of the water layer decreases. As pointed out
in teh previous section, median PWV between UT 01 hr and 11 hr (night) is 
about the same as between UT 12 hr and 21 hr (day), although the median 
value between UT 05 hr and 13 hr is significantly lower.
 
\begin{figure}[t]
\plotone{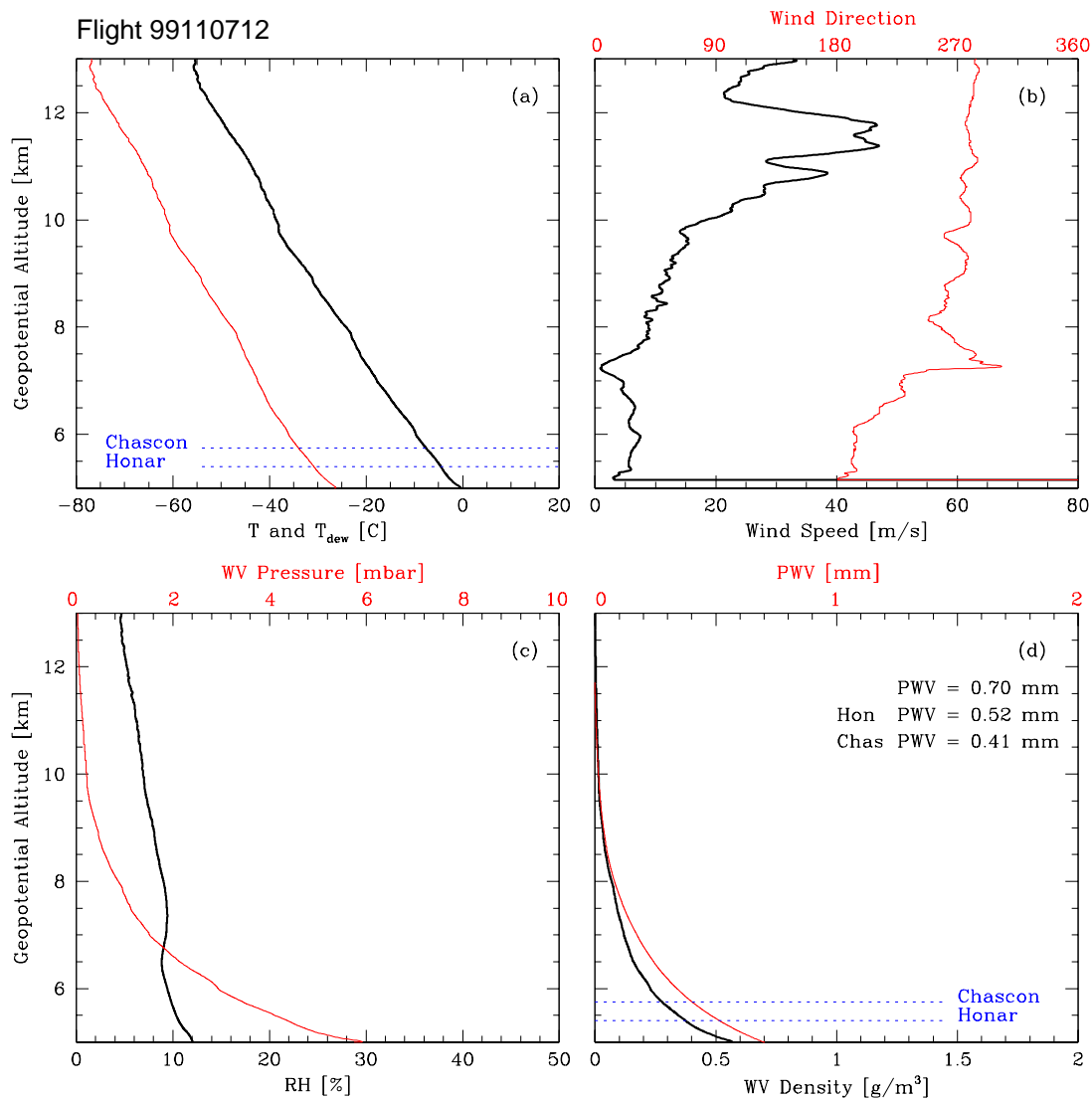}
\caption{Radiosonde atmospheric profiles. Panels display: at upper 
left, temperature 
(thick) and dewpoint temperature; at upper right, wind speed (thick) 
and direction; at lower left, relative humidity (thick) and water 
vapor pressure; at lower right, water vapor density (thick) and
PWV. The horizontal dotted lines are at the elevations of Cerro 
Honar (5400 m) and Cerro Chasc\' on (5750 m). Unusually smooth 
atmospheric conditions are sampled by this sonde flight of 7 November 
1999, UT 12 hr. No strong temperature inversions are seen in the lower 
atmosphere.   \label{99110712}}
\end{figure}

\begin{figure}[t]
\plotone{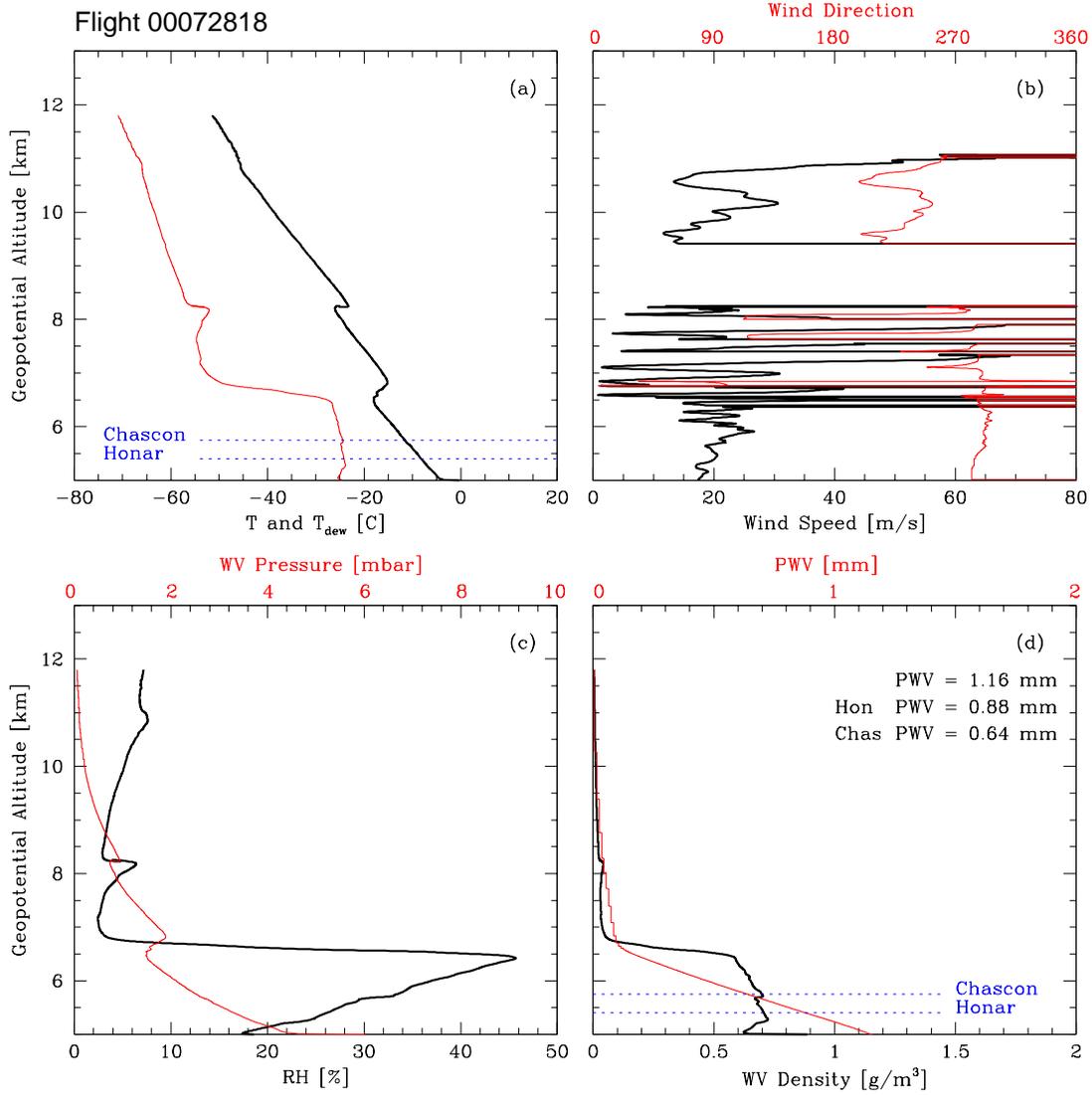}
\caption{Radiosonde data for a 28 July 2000 launch, UT 18 hr.
It shows temperature inversions near 6.6 and 8.2 km. Notice also 
the large increase in humidity below the 6.6 km inversion layer. 
Raggedness of wind data due to poor sonde tracking.
Panel description as in Figure \ref{99110712}. \label{00072818}}
\end{figure}

\begin{figure}[t]
\plotone{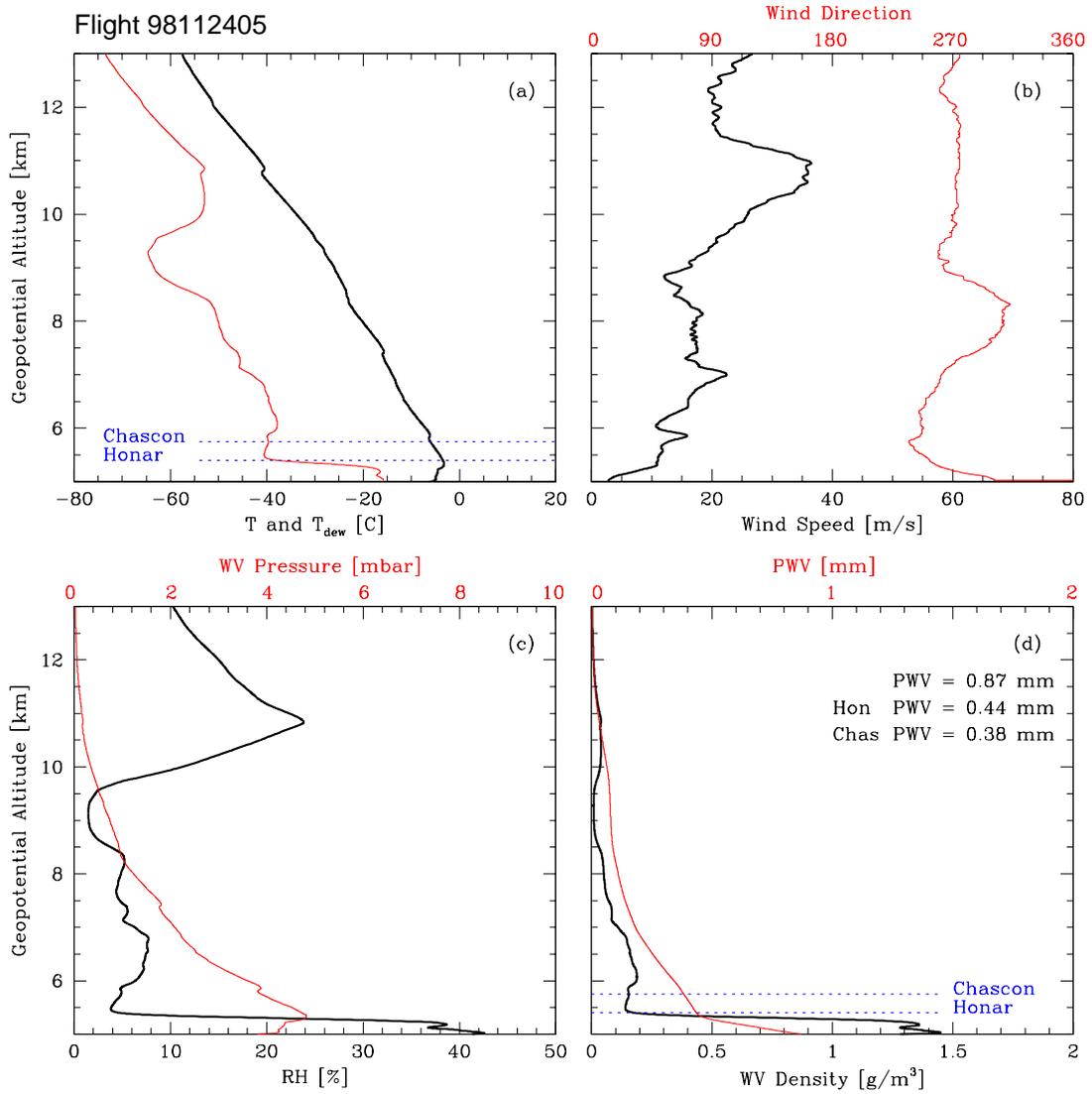}
\caption{Radiosonde data for a 24 November 1998 launch, UT 05 hr. 
It shows temperature inversion below 5.4 km. Panel description as 
in Figure \ref{99110712}.
\label{98112405}}
\end{figure}

\begin{figure}
\plotone{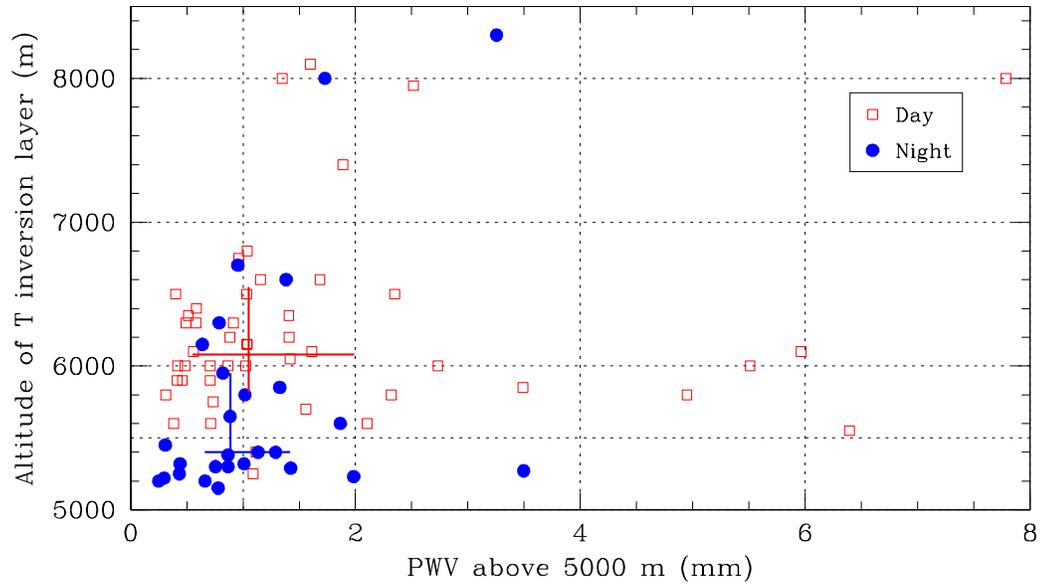}
\caption{Altitude of temperature inversion layers identified in
daytime (unfilled squares) and nighttime (filled circles) radiosonde
profiles. The large crosses are plotted at the median values of
elevation and PWV; the span of the crosses' arm extend from first
to third quartile. 
\label{t_inv}}
\end{figure}

On any given time, the departures of the water vapor distribution from 
an exponential can be quite severe.
Figures \ref{99110712}, \ref{00072818} ~and \ref{98112405} 
~display data for three different radiosonde launches. In panel (a) of each
figure, the air temperature and the dewpoint temperature are displayed;
in panel (b), wind speed and direction are shown; in panel (c), the relative
humidity and the water vapor pressure are given, while in panel (d) we have
the water vapor density and PWV. The temperature and water vapor profiles
of Figure \ref{99110712} ~are unusually smooth. The water vapor density 
distribution appears exponential, with a scaleheight of about 1420 m.
Profiles such as that tend to be the exception. More often, temperature
inversions are present, and the water vapor distribution is raggedly
uneven. In the case of Figure \ref{00072818}, for example, two prominent
temperature inversions take place near altitudes of 6600 and 8200 m.
The atmospheric water vapor lies mostly under the lower of the two inversion 
layers, and the vertical distribution of the water vapor density is far from 
exponential in shape. In Figure \ref{98112405}, an inversion layer is present 
near 5300 m, and half of all water vapor above the Chajnantor Plateau is 
packed in the lowest 400 m of atmosphere. Were circumstances such as 
that displayed in Figure  \ref{98112405} ~frequent, peaks that rise 
several hundred m above the plateau would be highly attractive candidate 
sites for infrared observations. We have reviewed all radiosonde profiles
and selected those for which temperature inversion layers can be
identified. In Figure \ref{t_inv}, we plot the altitude of the lowest
(in case that more than one is discernible) inversion visible in the 
temperature profile, separately for day and night radiosonde flights.
The fraction of temperature inversions taking place below 500 m above 
the plateau level is far larger during the night than during the day,
by an amply significant margin. The nightly decrease in the thickness 
of the water vapor layer appears to be related to the development of
temperature inversions close to the ground.

The occurrence and altitude of temperature inversions is of importance not only
for their impact on the PWV, and thus the infrared transparency, of a site, but
also for the quality of astronomical seeing. Evidence exists (Hufnagel 1978)  
that in those layers large values of the refractive
index structure constant $C_n^2$ occur. The seeing disk of stellar images, 
$\theta_{hpfw}\propto \int C_n^2 dz$, can thus be strongly affected. A site
above the inversion layer would hence yield better astronomical image quality.

\section {Atmospheric Transparency in the Infrared  \label{transp}}

\subsection {Atmospheric Transparency and Thermal Background}

\begin{figure}
\plotone{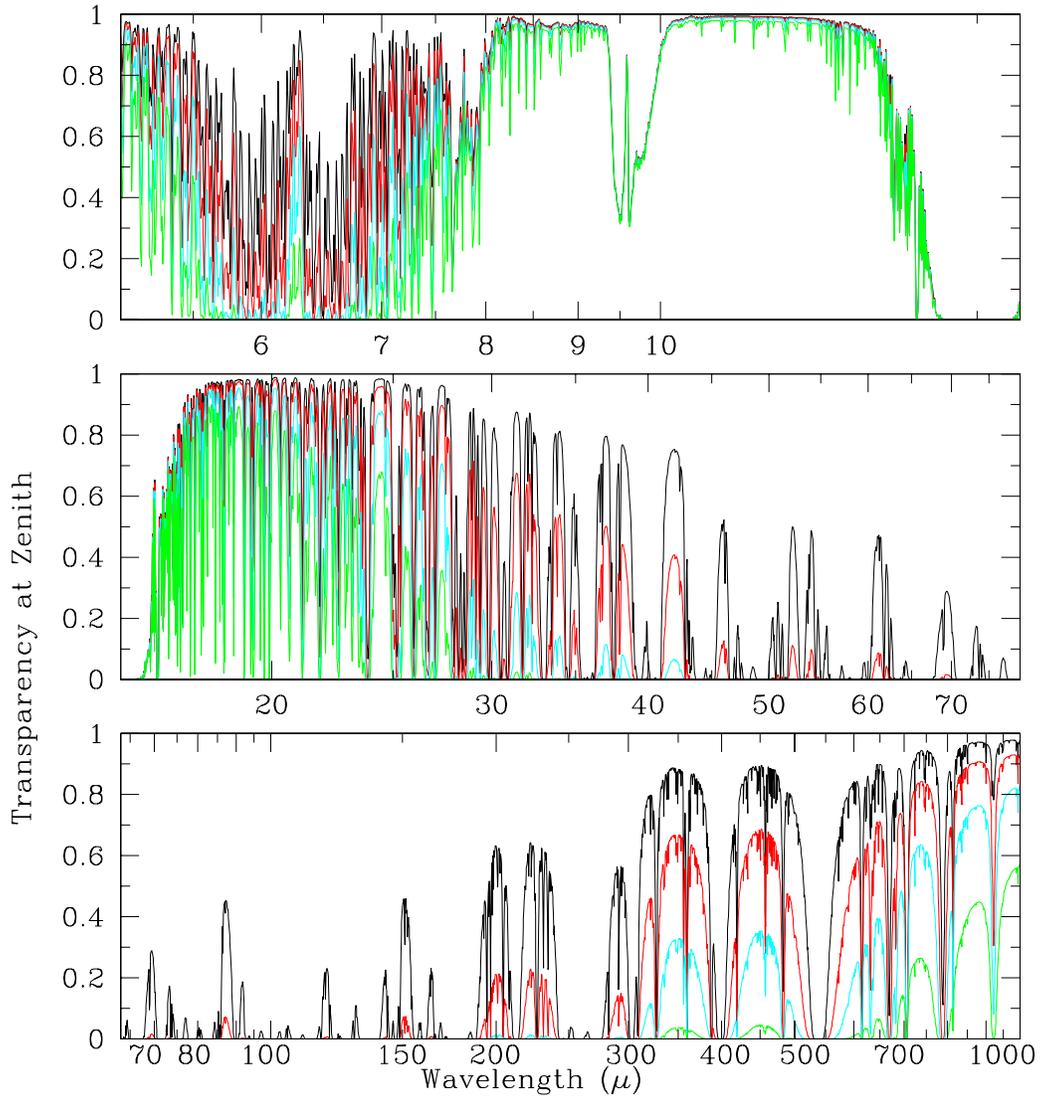}
\caption{Atmospheric transparency at zenith between 5 and 1000 \micron, 
for a site at an altitude of 5000 m and different H$_2$O column densities: 
the tracings, top to bottom, correspond to 0.1, 0.4, 1.0 and
3.0 mm of PWV. \label{trans}}
\end{figure}

Figure \ref{trans} ~displays curves of atmospheric transparency at zenith, 
computed for a site at an altitude of 5000 m and different column densities 
of water vapor above head, for the window between 5 \micron ~and  1 mm. In 
the spectral region between 50 and 200 \micron ~the atmosphere is quite
opaque from all ground sites: even if PWV were as low as 100 \micron,  
transparency would not reach 40\%. However, the window near 200 \micron ~may 
be of interest for observations of the very important spectral line of $C^+$.
For low values of PWV, numerous atmospheric windows appear longwards of 10 
\micron, notably near 20, 24, 32, 34, 38, 42 and 46 \micron.

The main deleterious effects of the atmosphere on astronomy consist in 
reduced transparency to cosmic photons from a given line of sight, seeing 
degradation of optical and infrared images and an increase of the diffuse 
background, against which the generally weak signals of cosmic sources need 
to be discriminated. The effective temperature of the most relevant atmospheric 
layers is between 200 and 300 K, so the atmospheric thermal emission peaks 
near 15  \micron. Near the centers of the absorption bands where the 
atmosphere is opaque, a maximum background flux is reached and, of course, no 
cosmic photons get through. In spectral regions of partial transparency, the 
atmospheric background radiation can be approximated by that of a blackbody 
attenuated by a factor which is the complement of the transparency. If opacity
is distributed over a wide range of elevations, this assumption may be flawed;
it is however a good approximation for the opacity produced by water vapor,
which is found very close to the ground, and is our main concern in this paper.
In the optical and near IR, the principal source of atmospheric background is 
the scattering of solar radiation, during the day. At night, however, airglow
in the near IR results from transitions between vibrational states of the OH$^-$
radical, discharging energy stored during daytime from solar radiation, through
the dissociation of ozone. Thus OH$^-$ airglow originates at heights of 70 to
90 km and affects all ground based locations, independently on altitude. 

In addition to the atmospheric background, astronomical observations need
to contend with cosmic backgrounds, such as that produced by the thermal
emission and the scattering of solar photons by interplanetary dust, the
interstellar dust emission, the cosmological microwave background radiation
and the thermal emission of the telescope and detectors themselves.

We next compute background levels for various atmospheric
conditions; in doing so we follow the approach of Thronson \etal (1995).

Let's first consider the ``near field'' thermal background.
In the atmospheric windows of interest the atmosphere is optically thin, so 
that the specific intensity of the emitted radiation can be approximated by
\be
I_\nu = \tau_\nu B_\nu = {2\tau_\nu (h\nu^3/c^2)\over exp(h\nu/kT) - 1}
\ee
where $B_\nu$ is Planck's function and $\tau_\nu$ the optical depth.
Thus the number of photons per unit time. originating with the thermal
background of temperature $T$ and striking the telescope within the solid
angle subtended by a point source $\Omega_{ps}$, is
\be
\bgm = I_\nu ~\atm ~\Omega_{ps} \Delta \nu/h\nu
\ee
where \at ~is the collecting area of the telescope and $\Delta \nu$ is the
detection bandwidth. Converting to photoelectrons per second and assuming
a detector gain of 1 
\be
\bgm_e = 2.22\times 10^{15} ~{\trsm~QE\over \resm}{{1-e^{-\tau}}\over 
\lambda_\mu[exp(1.44\times 10^4/\lambda_\mu T) -1]} ~~~~~{\rm [e~~s^{-1}]} 
\label{eq:B}
\ee
where \trs ~and $QE$ are respectively the system transmission and
quantum efficiency over the band, $\lambda_\mu$ the wavelength in \micron, and 
\res $= \nu/\Delta \nu = \lambda/\Delta \lambda$. Equation ~\ref{eq:B} 
is derived assuming that the telescope is diffraction limited at all wavelengths, 
so that $\Omega_{ps}$ is the solid angle subtended by the ring corresponding to 
the first diffraction null and $\atm \Omega_{ps}$ is then $3.7\lambda_\mu^2$. 
This assumption makes the background estimate independent on telescope aperture, 
allowing an intercomparison of different sites or atmospheric conditions. At the 
shorter wavelengths, at which the seeing angle is larger than the diffraction 
limit of the aperture, such an assumption is incorrect, and $\Omega_{ps}$ should 
be replaced with the solid angle which encircles 90\% of the light within the 
seeing disk.

Different sources of thermal emission yield an additive contribution to $\bgm_e$. 
For the atmospheric thermal background, we approximate $\tau$ with the complement 
to one  of the transparency as displayed in Figure \ref{trans}. For the thermal 
emission of the telescope, we replace $\tau$ in Eqn. \ref{eq:B} ~with a constant 
emissivity $\epsilon_{tel}$.  

As for the cosmic backgrounds, following Tokunaga (1998) and using Equation 
~\ref{eq:B}, we approximate the zodiacal emission of dust by that of a blackbody 
with $T=275$ K attenuated by $\epsilon = 7.1\times 10^{-8}$; the
sunlight scattered by the interplanetary particles by a blackbody of
$T=5800$ K and $\epsilon =3\times 10^{-14}$; the interstellar dust emission
by a blackbody of $T=17$ K and $\epsilon = 10^{-3}$; and the cosmic microwave
background by a blackbody with $T=2.73$ K and $\epsilon=1$.

\begin{figure}
\epsscale{1.0}
\plotone{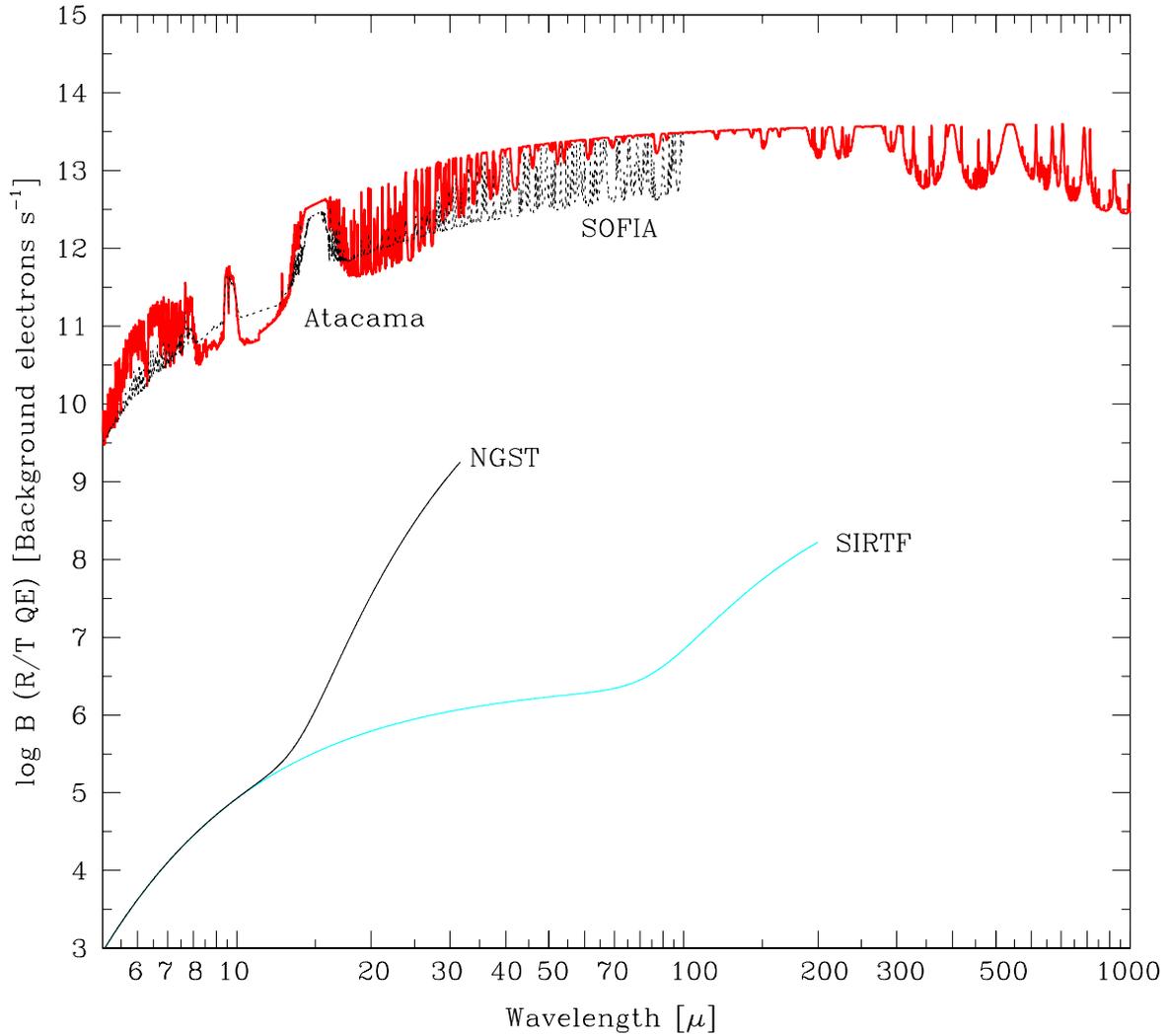}
\caption{The total backgrounds for Atacama (very best atmospheric
conditions with PWV=0.2 mm and seeing FWHM of 0.35" at 0.5 \micron 
~assumed) and several space and airborne telescopes are compared. 
Note that  that ${\trsm} ~QE/{\resm} \simeq 1$. See text for 
further details. \label{bgnd}}
\end{figure}

Figure \ref{bgnd} ~illustrates the background $\bgm_e$ for an Atacama telescope
with $\epsilon_{tel}=0.05$, estimated for the very best atmospheric conditions 
of $PWV=0.2$ mm and a seeing of 0.35" at 0.5 \micron (as shown in Table 1, PWV=0.2 
mm is the value for the lowest quartile at a high elevation site, and seeing
near 0.35" may have similar frequency at such a site; the chosen combination of
values may occur in 5--20\% of the time, depending on how correlated seeing and
PWV are; of eight reliable events of simultaneous radiosonde and seeing measurements, 
the four with  PWV$<1$ mm yield an average seeing of 0.51" at 0.5 \micron, while
the other four, with PWV$>1$ mm, have average seeing of 0.84"; more data is needed
to confirm this possible correlation). In the same figure, 
we also display estimates of the background for a space telescope passively cooled 
(e.g. the Next Generation Space Telescope, NGST), a cryogenically cooled space 
telescope (e.g. the Space Infrared Telescope Facility, SIRTF), and a stratospheric,
telescope airborne at 12 km altitude (above which PWV is assumed to be
10 \micron, $\epsilon_{tel} = 0.15$, and image quality of 2" at 1 \micron for SOFIA
(this may be an overestimate of image quality: while the optical quality of
the instrument is high, pointing jitter and seeing produced by the turbulent air
flow near the airplane may produce worse values). We have assumed a telescope 
temperature of 260 K for Atacama, 240 K for SOFIA, 60 K for NGST and 5.5 K for SIRTF.
The temperature of the emitting atmospheric layers was assumed to be 255 K
for Atacama and 230 K for SOFIA.
For simplicity, we assume 
${\trsm}\cdot QE\cdot G/{\resm} =1$, which overestimates the bandwidths that are 
possible in some of the atmospheric windows. Telescope and atmospheric thermal
backgrounds dominate in the case of all ground--based instruments, while
in the case of space instruments, even if not cryogenically cooled, the zodiacal 
emission dominates. Between 5 and 15 \micron, for example, the NGST 
will have a background about six orders of magnitude lower than 
that for an Atacama telescope. While airborne at high altitude, 
the SOFIA telescope itself will produce substantial thermal emission, and lower
image quality than can be achieved from the ground. Note that the background
in good Atacama conditions and for Sofia are comparable. SOFIA will, however, 
have access to atmospheric windows inaccessible from the ground.

\subsection {Comparison of Broad Band Sensitivities}

\begin{figure}
\plotone{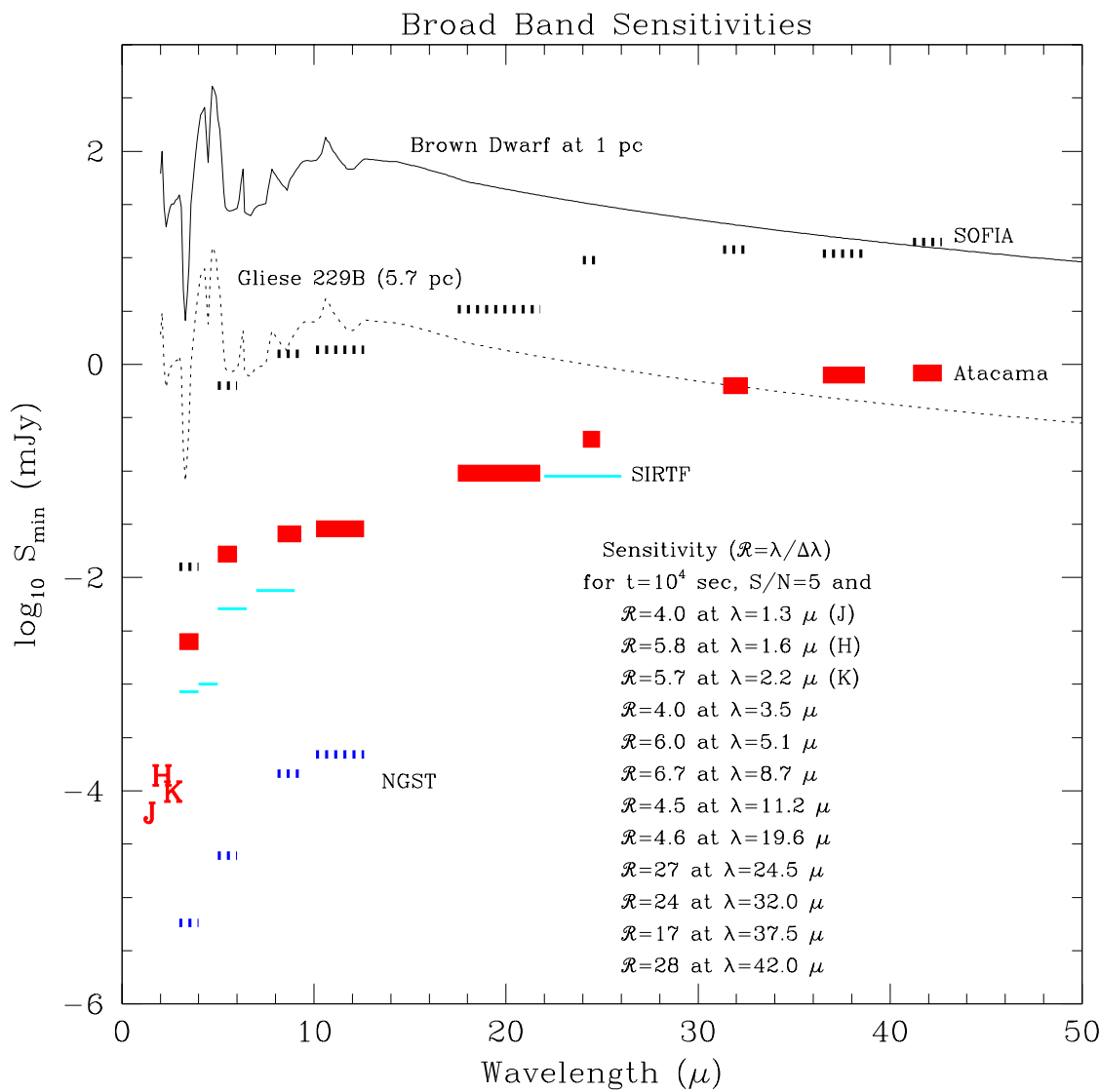}
\caption{Comparison of sensitivities for several proposed or future
telescopes, for a signal--to--noise ratio of 5 and an integration
of 10$^4$ s. The length of the horizontal bars identifying each 
waveband illustrates \res ~as tabulated in the inset. See text for
assumed telescope parameters. The 15 m Atacama telescope is assumed
seeing limited with image size of 0.35". The labels `J', `H' and
`K' refer to the Atacama sensitivities. The J band limit is
equivalent to about 26th magnitude. For reference, 
the spectral shape of a brown dwarf (Gliese 229B) has been plotted at
the assumed distances of 1 and 5.7 pc.
One mJy is $10^{-29}$ W m$^{-2}$ Hz$^{-1}$.
\label{fsens}}
\end{figure}

The signal--to--noise ratio \qu~ for a single integration of time $t$ can in 
general be expressed as (Thronson \etal 1995)

\be
\qum = {\sourcem~t\over [(\sourcem+\bgm_e+\darkm)t + R^2 + N_c t^2]^{1/2}} \label{eq:q}
\ee
where \source ~is the source signal, \dark ~is the detector dark current, $R$
its readout noise and $N_c$ is the signal produced by source confusion. A 
given signal, such as $S$, relates to the flux density via
\be
{\bf S} ({\rm mJy}) e^{-\tau} = 6.63\times 10^{-5}{{\resm} \over QE~{\trsm~\atm}}
~\sourcem \label{eq:F}
\ee
We adopt the following detector
performance characteristics for the 3--40 \micron ~ range: $QE\simeq 0.50$, 
$\darkm \simeq 10$ $e^- s^{-1}$ and $R\simeq 50$ electrons rms (Thronson 
\etal 1995; Stacey 2000; Brandl \& Pirger 2000), and source confusion does 
not become an issue shortwards of 
50 \micron, as we shall discuss in the next Section. It is thus safe to assume 
that in the near and mid--IR the thermal background dominates for broad band
observations, so that the minimum detectable signal, for an exposure of 
time $t$ achieving 
signal--to--noise \qu, according to Equation~\ref{eq:q} ~is
\be
\sourcem_{min} \simeq \qum~[\bgm_e/t]^{1/2}.  \label{eq:FS}
\ee
We use Equations \ref{eq:B} ~and \ref{eq:FS} ~to estimate limiting 
fluxes for point sources, assuming \trs$=0.4$ and, as indicated above, 
$QE=0.5$. The broadest bandwidths usable are, of course, limited by atmospheric 
absorption. Sensitivities are computed for $\qum =5$ and exposures
of $t=10^4$ sec. We underscore the fact that the chosen bands are those
accessible from the ground: SOFIA, SIRTF and NGST will access spectral regions
inaccessible from the ground. The results are plotted in Figure \ref{fsens}, for a
telescope at Atacama with a 15 m diameter, NGST of 6 m diameter, SIRTF with
a 0.85 m diameter and SOFIA with a diameter of 2.4 m. In the case of the
Atacama telescope, we have assumed that the image size is seeing--limited
at 0.35" at 0.5 \micron. Note that with an adaptive optics system of moderate 
to good efficiency, the near infrared sensitivity can improve by one order of 
magnitude, beyond the limit in Figure \ref{fsens}, provided that it does not
significantly increase the telescope emissivity. The implementation of an
adaptive secondary would thus be desirable and that of a multiconjugate 
system would raise more serious emissivity concerns. The point source sensitivity 
plotted in Figure \ref{fsens} ~for J band is equivalent to 26th magnitude; 
with an AO system that limit would approach 29th magnitude. 
The telescope labels are the same as for Figure \ref{bgnd}.
The SIRTF sensitivities have been obtained from the SIRTF website 
({\it http://ssc.ipac.caltech.edu/sirtf/Mission}) and links referred therein.
The NGST sensitivities have been plotted for the same bandpasses  and \res ~as 
for the Atacama telescope, although it should be clear that the whole near
IR spectrum will be accessible to NGST.
For comparison, we also plot the flux density curves of the brown dwarf 
Gliese 229B at the source's distance of 5.7 pc and at 1 pc (Matthews \etal 1996).


The thermal background does not always constitute the main limitation
at submillimeter wavelengths. Source confusion can become important. The sky 
source density at far IR and submm wavelengths is relatively uncertain, but
it is unlikely to be a concern even for large integrations ($\sim 10^4$ s)
for large aperture telescopes (diameter greater than 10 m). Assuming
${\trsm}=0.4$, $QE=0.5$ (a quantum efficiency which can be approached by 
detectors with modern bolometers) and optimal atmospheric conditions as can 
be found in Atacama (PWV=0.2 mm), an observation with $\qu = 5$ and $t=10^4$ s
can reach a sensitivity on the order of 1 mJy at all the atmospheric windows
between 300 \micron ~and 1 mm, and about 3 mJy for those between 200 and 300
\micron.

\subsection {Comparison of Spectroscopic Sensitivities}

Suppose the flux density contributed at a given
frequency by a spectral line is {\it S}. The integrated flux over the
whole line is $\int S d\lambda$. If the spectral line is unresolved,
the flux associated with the line is $S\Delta \lambda = S\lambda/\resm$,
where {\it S} is now the mean line flux over the spectral channel which 
contains the line. If we 
measure the integrated flux in W m$^{-2}$, then the signal $S$ in electrons
per second can be converted to
\be
F{\rm (W~m}^{-2}) = 2.0\times 10^{-19} {S\over \lambda_\mu~A~QE~\trsm}
\label{eq:F}
\ee
In spectroscopic observations, the high \res ~reduces the impact of the
high thermal background, as illustrated by
Equation ~\ref{eq:B}. For observations with \res $=10^5$, for example,
the thermal background contribution,
which is inversely proportional to \res, will be several orders of
magnitude lower than in the case of photometric observations. Consider 
Equation ~\ref{eq:q}: $\bgm_e$ can become comparable with or smaller than
\dark, in which case the observation will be limited by detector noise. For
values of $\darkm \simeq 10$ e$^-$ s$^{-1}$ and $R\simeq 50$ electrons,
the background can easily become negligible with respect to \dark ~in
space--based instruments. The readout noise on the other hand is likely
not to play a role if exposures exceed a few hundred seconds, because
then $R^2 < \darkm t$. We will assume here that $R^2$ and the source 
confusion term in the denominator of Equation~\ref{eq:q} ~are negligible.

\begin{figure}
\plotone{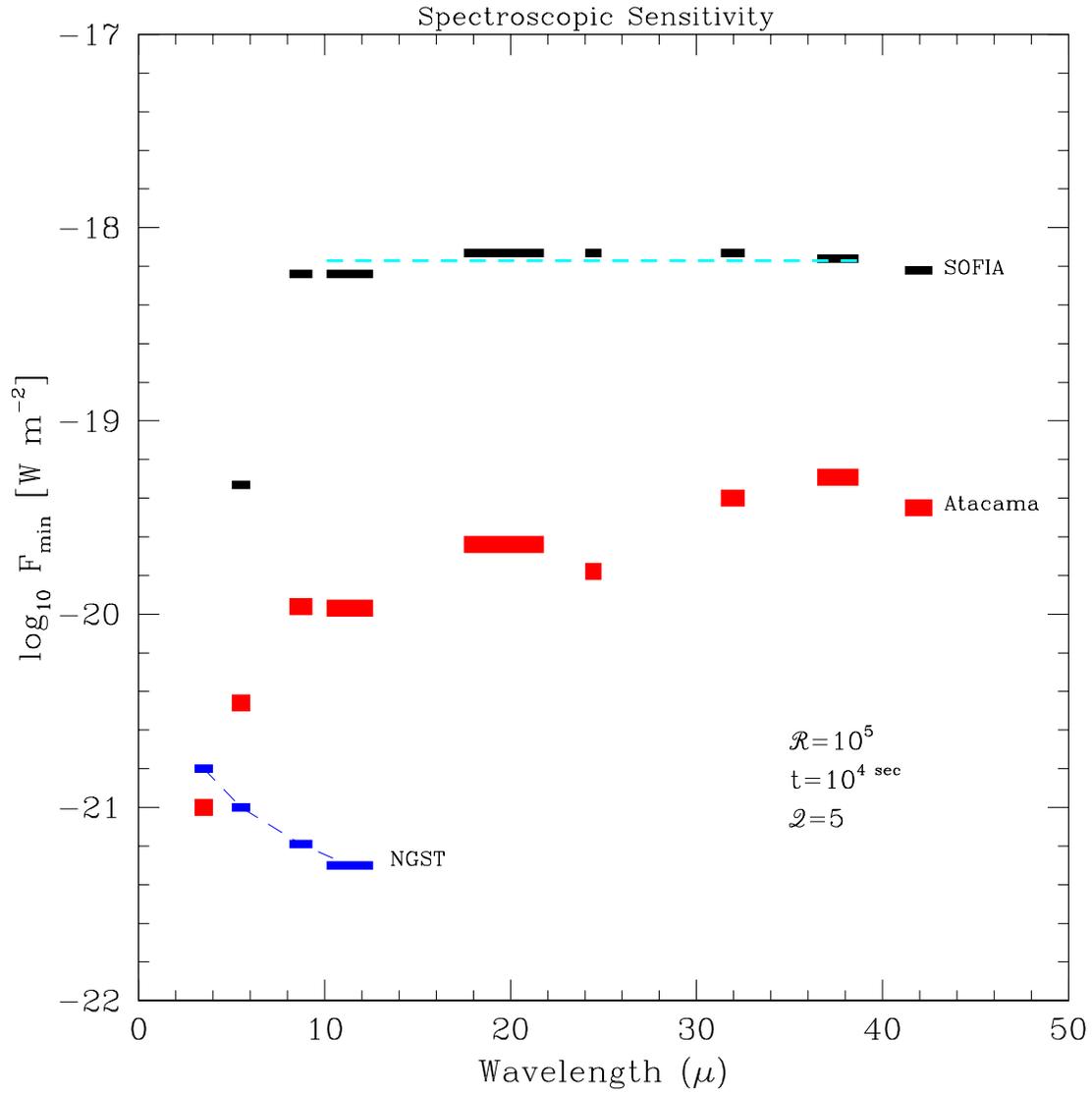}
\caption{Spectral sensitivities for several proposed telescopes;
for Atacama, calculations correspond to \res$=10^5$, $\qum=5$ and 
$t=10^4$ sec. The spectral sensitivity of SIRTF, at its highest resolution 
of 600, is indicated for comparison by a horizontal dashed line.
\label{fspec1}}
\end{figure}

The minimum detectable integrated line flux for an integration time $t$, with 
a signal--to--noise ratio \qu, can be written, in analogy to 
Equation~\ref{eq:FS}, as
\be
F_{min}{\rm (W~m}^{-2}) = 2.0\times 10^{-19}
{\qum \over \lambda_\mu~A~QE~\trsm}  
~\Bigl[{\bgm'\over t}\Bigr]^{1/2}
\label{eq:fmin}
\ee
where $\bgm'$ is $\bgm_e + \darkm$. Note that, for a space--based instrument,
near 5 \micron ~$\bgm_e \simeq 5\times 10^{-3}$ e$^{-}$ s$^{-1}$ for
\res$ = 10^5$, so $\bgm_e \ll \darkm$, and the instrument is detector--limited.
In  Figure \ref{fspec1}, we show sensitivity limits computed using
Equation ~\ref{eq:fmin}, for \res$=10^5$, \dark=10, $QE=0.5$, \trs=0.4
and \qu=5, except for SIRTF, for which \res ~is only 600.

\section {Conclusions \label{wvs}}

The water vapor distribution above the Chajnantor plateau appears to have
the following properties:

\begin{itemize}
\item The median PWV above the plateau at 5000 m  is $\sim 1.2$ mm. Variations
from year to year can be as large as 50\% of that value.
\item Seasonal variations in PWV are also large: median values for the 8 months 
from April to November are 30\% lower than the yearly median, while median values 
for the Summer months may be more than twice as large.
\item PWV varies by about $\pm 20$\% with the daily cycle, with a phase lag 
about 4 hours behind that of sunlight: minimum PWV occurs between midnight 
and noon, and maximum occurs at sunset.
\item The distribution of the water vapor of the median atmosphere over
Chajnantor is well aproximated by an exponential, with a scaleheight of 1.13 km. 
At any given time, however, the water vapor distribution can depart 
from an exponential shape, more dramatically when temperature inversions
occur.
\item The thickness of the water vapor layer appears to vary in phase with 
the solar illumination cycle by about $\pm 25$\%; minimum is near local midnight.
\end{itemize}

The implications of these effects for summits in the vicinity of the
Chajnantor Plateau are:

\begin{itemize}
\item The median PWV above 5400 m elevation drops by one third with respect 
to the value measured at the plateau; above an elevation of 5750 m it drops 
by  a factor of 2. 
\item Median PWV in winter nights at a summit near 5400 m, such as Cerro 
Honar, may be as low as 0.5--0.6 mm, if conditions near the summit approach 
those of the free atmosphere. The lowest quartile of PWV may approximate 
0.35--0.40 mm.
\item Median PWV in winter nights at a summit near 5700 m, such as Cerro 
Chasc\' on may approximate 0.40 mm, and the lowest quartile near 0.20--0.25 mm.
\item Low elevation temperature inversions are more likely during nighttime.
During those episodes, much of the PWV is trapped below the inversion layer, 
and mountain peaks are offered a nearly dry atmosphere, possibly endowed with 
high quality seeing.
\end{itemize}

The conclusions listed above rely on still relatively sparse data from 108
radiosondes and should be considered as preliminary. However, these results
suggest that exceptional possibilities for ground--based IR and submm 
astronomical observations exist in the Llano de Chajnantor
region. The combination of low water vapor content and high quality seeing 
allow for low atmospheric background in the near and mid--IR. At a telescope 
on a summit in the vicinity of the Chajnantor plateau, numerous 
atmospheric windows would appear in the mid IR, up to about 50 \micron. In the
far IR and submillimeter regime, 
the 350 \micron, 450 \micron, 600 \micron, 750 \micron ~and 870
\micron ~windows reach exceptional transparency, while two useful windows 
appear near 200 \micron. Broad band observations with a 15 m class telescope
at such a site would be close in sensitivity to those made with SIRTF in the
mid--IR, offering a superb synergy match for follow--up observations of the 
SIRTF surveys. The near and mid--IR performance of such a ground--based telescope 
in high resolution spectroscopic mode would be exceptional and could only be 
exceeded by that of a space telescope of comparable aperture. The relative
ease and cost--effectiveness of operation in the Atacama advises that 
serious attention be given
to the Atacama sites for the installation of the next generation of large
infrared telescopes.

{\bf Acknowledgements:} The support and assistance of Joe Veverka, 
Yervant Terzian, Bob Richardson, Jennifer Yu and Bryan Isacks of Cornell U.; 
Martha Haynes of Cornell U. and Associated Universitites, Inc.; Robert L. Brown, 
Eduardo Hardy and Geraldo Valladares of NRAO; the use of radiosonde data obtained 
through a collaboration with NRAO (Simon Radford and Bryan Butler), ESO (Angel
Otarola) and SAO (Ray Blundell and Scott Paine); access to 225 MHz radiometry
data of NRAO (Simon Radford); don Tom\' as Poblete Alay and the 
staff of {\it La Casa de Don Tom\' as} are thankfully acknowledged. 
This study was made possible by a grant of the Provost's Office of Cornell 
University and the National Science Foundation grant AST--9910136.


\begin{deluxetable}{ccccr}
\tablewidth{0pt}
\tablenum{1}
\tablecaption{PWV (mm) Quartiles for 108 Radiosondes. \label{pwvtab}}
\tablehead{
\colhead{Quartile} & \colhead{5000 m} & \colhead{5400 m} & \colhead{5750 m} & N
}
\startdata
{\bf All}   &  --- & ---  & ---  & 108 \nl
25\%  & 0.71 & 0.40 & 0.27 & \nl
50\%  & 1.04 & 0.72 & 0.49 & \nl
75\%  & 1.75 & 1.28 & 0.92 & \nl
{\bf Day}   &  --- & ---  & ---  &  65 \nl
25\%  & 0.58 & 0.40 & 0.32 & \nl
50\%  & 1.04 & 0.82 & 0.54 & \nl
75\%  & 1.75 & 1.33 & 1.00 & \nl
{\bf Night} &  --- & ---  & ---  &  30 \nl
25\%  & 0.76 & 0.37 & 0.21 & \nl
50\%  & 1.00 & 0.57 & 0.42 & \nl
75\%  & 1.42 & 1.05 & 0.68 & \nl
{\bf UT 05--13} & --- & --- & ---&  32 \nl
25\%  & 0.46 & 0.26 & 0.20 & \nl
50\%  & 0.85 & 0.53 & 0.36 & \nl
75\%  & 1.23 & 0.86 & 0.63 & \nl
\enddata
\end{deluxetable}

\newpage

\end{document}